\documentstyle[12pt,psfig]{article}

\newcommand{\beq}[1]{\begin{equation}\label{#1}}
\newcommand{\eeq}{\end{equation}}
\newcommand{\beqar}[1]{\begin{eqnarray}\label{#1}}
\newcommand{\eeqar}{\end{eqnarray}}

\newcommand{\Mfunction}[1]{{\rm #1}}

\newcommand{\bra}[1]{\big< #1 \big|}
\newcommand{\ket}[1]{\big| #1 \big>}
\newcommand{\nn}{\nonumber}
\newcommand{\g}{{\rm g}} 
\newcommand{\f}{{\rm f}} 
\newcommand{\dd}{{\rm d}} 
\newcommand{\Gl}[1]{Eq.~(\ref{#1})}
\newcommand{\Ab}[1]{Fig.~\ref{#1}}

\newcommand{\al}{\alpha}
\newcommand{\be}{\beta}  
\newcommand{\ep}{\varepsilon}
\newcommand{\ga}{\gamma}

\newcommand{\et}{\eta}
\newcommand{\la}{\lambda}

\newcommand{\si}{\sigma}

\newcommand{\La}{\Lambda}

\begin{document}
\vspace*{-2cm}
\hfill DFTT 13/98, TPR-98-13, TUM/T39-98-8
\vspace{4cm}
\begin{center}
{\Large \bf Renormalon Model Predictions for Power-Corrections to \\[2mm]
Flavour Singlet Deep Inelastic Structure Functions}
\vspace{1cm}
\end{center}
\centerline{E. Stein$^a$, 
M.~Maul$^b$, L.\ Mankiewicz\footnote{On leave of absence from
N. Copernicus Astronomical Center, Polish Academy of Science, ul. Bartycka 18,
PL--00-716 Warsaw (Poland)}
$^c$, A.~Sch\"afer$^b$}
\vspace{1 cm}

\centerline{$^a${\em  INFN Sezione di Torino, 
Via P.~Giuria 1, I-10125 Torino, Italy}}  

\centerline{$^b${\em  Naturwissenschaftliche Fakult\"at II, 
Universit\"at Regensburg, 
D-93040 Regensburg, Germany}} 

\centerline{$^c${\em  Physik Department, 
Technische Universit\"{a}t M\"{u}nchen,}} 
\centerline{\em D-85747 Garching, Germany}

\vspace{1 cm}

\noindent{\bf Abstract:}
We analyze power corrections to flavour singlet deep inelastic
scattering structure functions in the framework of the infrared
renormalon model. Our calculations, together with previous 
results for the non-singlet contribution, allow to model the 
$x$-dependence of higher twist corrections to 
$F_2, F_L$ and $g_1$ in the whole $x$ domain.
\vspace*{\fill}
\eject
\newpage

\section{Introduction}
Since the historical experiments of the SLAC-MIT group \cite{hist1} revealed
the partonic character of the nucleon's constituents, deep-inelastic
scattering (DIS) has been one of the most important tools to increase our
understanding of the inner structure of the nucleon.

The first experiments established that nucleons can be viewed 
as consisting of point-like constituents,
which later were identified as quarks and gluons.
The observed scaling behavior of the lepton-nucleon scattering
cross section was interpreted as incoherent scattering of
the probing lepton with the partonic constituents of the nucleon.

Since the advent of quantum chromodynamics (QCD) considerable
substance has been added to this simple picture.
The scaling behavior has found an explanation through
the asymptotic freedom of QCD. 
Asymptotic freedom combined with the factorization property resulted in a
systematic expansion of the DIS cross section
in terms of the coupling constant $\al_S(Q^2)$, 
evaluated at the scale set by the 
virtuality of the hard photon.
The leading order term in this expansion accounts for
the simple parton model result, while higher order 
corrections modify the scaling behavior, resulting in a logarithmical
scale dependence.

The QCD explanation of logarithmic scaling violations in DIS, as observed in a
later set of experiments \cite{hist2}, is considered as one of the most
important successes of the theory.

Since then, the experimental precision of DIS measurements has
reached remarkable accuracy over a wide range in $Q^2$. Using these data it has
been in principle possible to study not only the logarithmic 
corrections to scaling, but also to extract higher twist terms, i.e., power
suppressed corrections that fall off like powers of $1/Q^2$ \cite{VM92}.

From the theoretical side, the best tool to analyze such power corrections
is the framework of the operator product expansion (OPE) \cite{Pol73}. Twist-4
corrections to DIS have been studied systematically e.g. in \cite{Jaff82}.
Kinematical 
corrections arising from the non-vanishing mass $M$ of the target hadron fall
off like powers of $M^2/Q^2$ and can be attributed to power suppressed
contributions of twist-2 operators. They can be taken into account exactly by
introducing the so-called Nachtmann scaling variable \cite{Na??}.  

The other set arises due to the higher twist operators that are
sensitive to multi-parton correlations in the target. While estimates
of twist-2 matrix elements from lattice QCD are already 
available \cite{Goe95}, reliable estimates of higher twist 
matrix elements are not yet feasible.

In particular, calculation of higher twist operators 
with quantum numbers which do not prohibit mixing
with lower twist operators is a severe theoretical problem that has not
yet been solved \cite{Mar96}. 

The problem originates from the fact that twist-4 operators, in addition to
the usual logarithmic scale dependence due to their renormalization, may
exhibit quadratic UV divergences. Recall that the matrix element of a twist-2
operator is a dimensionless number. A related twist-4 operator of the same
spin and quantum numbers has therefore a matrix element of dimension 2. 
Then, radiative corrections result in a contribution of the
form of a square of the UV cut-off multiplied by the lower order matrix
element of the twist-2 operator. Such a mixing makes the definition of
the twist-4 contribution ambiguous. In the OPE of a physical quantity,
like DIS structure 
functions, this ambiguity always cancels against the corresponding ambiguity
in the definition of the twist-2 contribution. The latter arises because of the
asymptotic character of the QCD perturbation series \cite{ren1,ren2}. Hence,
the sum of twist-2 
and twist-4 contributions is unambiguous up to order $1/Q^2$, 
provided that both are calculated
within the same regularization scheme.

Recently, this subtle relation between twist-2 and twist-4 contributions has
motivated a phenomenological hypothesis \cite{BBM97} stating that the main
contributions to matrix elements of twist-4 operators are proportional to
their quadratically divergent parts\footnote{
  If quantum numbers prohibit mixing of the twist-4 operator with a lower
  dimensional twist-2 operator, the former exhibits, of course, no
  quadratic divergences of its matrix elements. It would be very
  interesting to find a set of experimentally accessible observables which
  would allow to extract a power correction which cannot be interpreted as a
  UV dominated twist-4 matrix element.}. 
In processes which cannot be analysed
in terms of OPE one extracts information about power corrections directly from
the large-order behavior of the corresponding perturbative series. Power
suppressed corrections to various observables, like event shape variables in
DIS \cite{DMW96} and $e^+\, e^-$ annihilation, 
as well as DIS structure functions
\cite{SMMS,DW96,MMMSS} have been shown to follow the behavior 
predicted by the UV-dominance hypothesis 
(for recent reviews see \cite{We97,Za98}).
Hence, with all reservations, it can be considered as a useful
phenomenological tool for estimating power suppressed contributions.

Comparison with the existing experimental data on leading power corrections to
the $F_2(x,Q^2)$ structure function \cite{VM92} has shown that the non-singlet
IR renormalon calculation describes the data on proton and deuteron structure
functions very well in the region of Bjorken-$x > 0.25$ but there is a
systematic discrepancy in the region $x< 0.25$ \cite{MSSM}.  In the present
paper we extend the existing analysis of power corrections to flavour 
non-singlet structure functions \cite{SMMS,DW96,MMMSS} 
to the quark pure flavour singlet case. We
follow the idea to trace twist-4 operators by calculating their UV divergent
part in the renormalon approach.
As expected, the resulting correction turns
out to be much smaller than the non-singlet one in the large-$x$
domain, but it is substantial in the small-$x$ region. It is encouraging to 
see that our predictions agree with the tendency seen in the data points 
below $x<0.2$ \cite{VM92}. 

Extending the existing measurements deeper into the small-$x$ --
small-$Q^2$ domain of DIS one ultimately enters a 
transition region between
interactions of hard and (almost) real photons with a nucleon. The twist
expansion provides a tool to approach this transition from the large-$Q^2$
side.  
The importance of twist-4 corrections to $F_2(x,Q^2)$ in the small $x$ domain
has been realized a long time ago \cite{Bar93}. The four-gluon operator was 
identified as a potential source of large twist-4 corrections
\cite{Bar93}, and its leading log anomalous dimension was 
calculated \cite{Bar93,Lev92}. 
Recently, a model analysis of twist-4 contributions has been
performed in the diagrammatic language \cite{MarRys98}. 
We emphasize that the renormalon 
analysis of twist-4 contributions in the pure flavour singlet sector, 
presented in the present 
paper, cannot substitute a non-perturbative QCD calculation of these
corrections. Nevertheless, as we shall discuss in
details below,
certain interesting features arise which can be confronted with existing
data. 

Recently, new data have been published \cite{H1,ZEUS} which extends a previous
NMC analysis \cite{NMC95,NMC} further into the small-$x$ -- small-$Q^2$ region.
 Down to $Q^2$
of the order of 1 GeV$^2$ the data can be well described by a set of
radiatively 
generated twist-2 parton distributions \cite{GRV}. On the other hand, even if
the precise form of its $x$-dependence is 
still subject to debate, there is certainly no reason to assume that twist-4
corrections are small and that they can be neglected. In our opinion, it would
be 
interesting to have the same data reanalyzed using a model for the twist-4
contribution e.g., in the form derived in the present paper or taken from
Ref.~\cite{MarRys98}. 

In the following we will investigate power corrections 
to the three structure functions $F_2$, $F_L$ and $g_1$
that appear in the well known decomposition of the 
hadronic scattering tensor of deep inelastic lepton nucleon scattering
\cite{Bar78}:
\beqar{wmunu} W_{\mu\nu}(p,q) &=& \frac{1}{2\pi}\,\int d^4z e^{i q z }
\bra{pS} J_\mu(z) J_\nu(0) \ket{pS}
\nn \\
&=& \left(\g_{\mu\nu} - \frac{q_\mu q_\nu}{q^2} \right) \frac{1}{2x}
F_L(x,Q^2) - \left(\g_{\mu\nu} + p_\mu p_\nu \frac{q^2}{(p\cdot q)^2} -
  \frac{p_\mu q_\nu + p_\nu q_\mu}{p \cdot q} \right)\frac{1}{2 x} F_2(x,Q^2)
\nn \\
&& - i \epsilon_{\mu\nu\la\si}\frac{q^\la S^\si}{p \cdot q} g_1(x,Q^2) 
\quad.
\eeqar
Here $J_\mu$ is the electromagnetic quark current, $x = Q^2/(2p \cdot q)$ and
$q^2 = -Q^2$. The nucleon state $\ket{pS}$ has momentum $p$ and spin $S$ with
$S^2=-M^2$, $M$ being the nucleon mass.  The convention for the
$\epsilon$-tensor has been taken from \cite{IZ}. We have neglected terms
arising due 
to weak interactions as well as the second spin
dependent structure function $g_2(x,Q^2)$ which is suppressed kinematically.

Our presentation is organized as follows. In section 2 we present the basic
assumptions and definitions of the renormalon model.  In section 3 we describe
our calculation, which leads to the estimate of the $x$-dependence of the
twist-4 corrections to DIS in the flavour singlet channel. Explicit
calculation is performed for the pure-singlet quark contribution, and
subsequently this result is used to model the $x$-dependence of the
corresponding gluon contribution. Section 4 is devoted to phenomenological
discussion 
of our results. We discuss the renormalon model predictions for $1/Q^2$
corrections to unpolarized and polarized nucleon structure functions with a
particular emphasis on the small $x$-dependence of these corrections. A
possible data fitting procedure is discussed in the summary. Finally,
our appendix contains the list of all renormalon model formulae derived in the
present paper.

\section{Nucleon structure functions beyond the leading twist}

According to the OPE, hadronic structure functions $F_i,
i=L,2$ in (\ref{wmunu}) can be decomposed up to ${\cal O} ( \frac{1}{Q^4})$ accuracy as
\beq{eq:def_twist_decomp}
F_i( x,Q^2) = F_i^{\rm tw-2}(x,Q^2) + \frac{1}{Q^2} h_i^{\rm TMC}(x,Q^2)
                                + \frac{1}{Q^2} h_i(x,Q^2) 
                                + {\cal O} ( \frac{1}{Q^4})\quad,
\eeq
where $F_i^{\rm tw-2}$ describes the leading twist-2 contribution.  $h_i^{\rm
  TMC}$ describes the target mass corrections which are directly related to
twist-2 matrix elements \cite{FG80}. It is  $h_i$ which contains the genuine twist-4
contribution and is sensitive to multi-parton correlations within
the hadron. The goal of this paper is to provide a phenomenological model for
  coefficients $h_i(x,Q^2)$ in the flavour-singlet sector.

In the QCD-improved parton model the twist-2 contribution to a deep inelastic
structure function can be represented as a convolution of process independent,
universal parton densities with perturbative coefficient functions.  For a
general number 
of flavours $n_f$ the corresponding formula reads:
\beqar{sfunction}
F_i^{\rm tw-2}(x,Q^2) &=& x \int_x^1 \, \frac{\dd z}{z}
\bigg\{\big(\frac{1}{n_f}\sum_{k=1}^{n_f} e_k^2 \big) 
\bigg[\Sigma(x/z,\mu^2)\, C_{i,q}^S(z,Q^2/\mu^2)
\nn \\
&&+ G(x/z,\mu^2) \, C_{i,G}(z,Q^2/\mu^2)\bigg] 
+  \Delta(x/z,\mu^2) \, C_{i,q}^{NS}(z,Q^2/\mu^2)\bigg\} \quad.
\nn \\
\eeqar
Here $G(x,\mu^2)$ denotes the gluon density, $\Sigma(x,\mu^2)$,
$\Delta(x,\mu^2)$ stand for the singlet (S) and non-singlet (NS)
combinations of quark densities, and $C_{i,G}$, $C_{i,q}^S$, and
$C_{i,q}^{NS}$ represent Wilson coefficients in the corresponding channels.
$F_i$ represents $F_L$ or $F_2$.
The factorization scale is denoted by $\mu^2$. 

The flavour singlet combination of the quark densities is
defined as
\beq{sdens}
\Sigma(z,\mu^2) = \sum_{i=1}^{n_f} \left(f_i(z,\mu^2) + \bar{f}_i(z,\mu^2)\right)
\quad,
\eeq
where $f_i$ and $\bar{f}_i$ stand for quark and antiquark densities
of species $i$. 
The non-singlet combination is given by
\beq{nsdens}
\Delta(z,\mu^2) = \sum_{i=1}^{n_f}\left(e_i^2 - \frac{1}{n_f}
\sum_{k=1}^{n_f} e_k^2\right) \left(f_i(z,\mu^2) + \bar{f}_i(z,\mu^2)\right)
\quad.
\eeq
The charge of the quarks is denoted by $e_i$.
A decomposition completely equivalent 
to \Gl{sfunction} can be written for the polarized structure function 
$g_1$ with 
polarized quark and gluon densities $\Delta\Sigma$ and $\Delta G$, and 
coefficient functions $\Delta C_q$ and $\Delta C_g$, respectively.

The parton densities are pure twist-2 quantities. Taking moments of
structure functions one obtains the OPE relation
\beqar{ope}
M_{k,N}(Q^2) &=& \int\limits_0^1 \dd x \; x^{N-2} F_k^{\rm tw-2}(x,Q^2) \nn \\
&=& \tilde{C}_{k,N}\left(\frac{Q^2}{\mu^2}, a_s\right)
\left[A_N(\mu^2)\right] \; .
\eeqar
Here $a_s$ stands for 
\beq{alphas}
a_s = \frac{g^2}{16 \pi^2} = \frac{\alpha_s}{4 \pi} \quad,
\eeq
and the $A_N$ are the matrix elements of the spin-N twist-2 operators which
determine the parton densities in \Gl{sfunction}. $F_k$ can be $F_L$, $F_2$ or
$x g_1$.The coefficient functions can be calculated order by order in 
perturbation theory. At present, Wilson coefficients are available up to
the second order \cite{Zijl92} and in some cases, for the lowest
moments, up to the third order in $a_s$ \cite{Lar96}.

Matrix elements of twist-4 operators \cite{Jaff82} 
contribute power corrections to the
simple partonic picture in \Gl{sfunction}. According to the UV dominance
hypothesis they can be considered proportional to their quadratically
divergent part or, equivalently, to the uncertainty of the perturbation series
in the definition of the twist-2 contribution. Commonly, the divergent series
of radiative corrections is regarded as an asymptotic series and defined by
its Borel integral. The actual calculations are done in the limit of 
large $n_f$,
which allows to resum the fermion bubble-chain to all orders yielding the
coefficient of the $a_s^n n_f^{n-1}$ - term exactly. Subsequently, it is
converted into the exact coefficient of the $a_s^n \beta_0^{n-1}$ - term
by the substitution $n_f \to n_f - 33/2 = - \frac{3}{2} \beta_0$, known as the
'Naive Non-Abelianization' (NNA)
\cite{Bro93,Ben93}. 
The asymptotic character of the resulting perturbative series
leads to resummation ambiguities - a singularity in the Borel integral
destroys the unambiguous reconstruction of the series and shows up as a
factorial divergence of the coefficients of the perturbative expansion.  The
general uncertainty in the perturbative prediction can be estimated to be of
the order of the minimal term in the expansion, or by taking the imaginary
part (divided by $\pi$) of the Borel integral. Both procedures lead to
resummation ambiguities of order  $\left(\Lambda^2/Q^2\right)^r$,
with $r=1$ for the leading IR renormalon. The resulting model for the twist-4
contributions $h_i$, see \Gl{eq:def_twist_decomp}, has the form of a Mellin
convolution :
\beqar{uvpart}
h_{i,G}(z,\mu^2) & = & (\La_{i,g})^2 \, x \int_x^1 \, \frac{\dd z}{z}\, 
A_{i,G}^{(2)}(z,\log((\La_{i,G})^2/\mu^2)) \, G(x/z,\mu^2)
\nn \\ 
h_{i,q}^{S}(z,\mu^2) & = & (\La_{i,q}^{S})^2 \, x \int_x^1 \, 
\frac{\dd z}{z}\, A_{i,q}^{S\;(2)}(z,\log((\La_{i,q}^{S})^2/\mu^2)) \, 
\Sigma(x/z,\mu^2)
\nn \\
h_{i,q}^{NS}(z,\mu^2) & = & (\La_{i,q}^{NS})^2 \, 
x \int_x^1 \, \frac{\dd z}{z}\,
A_{i,q}^{NS\;(2)}(z,\log((\La_{i,q}^{NS})^2/\mu^2)) \,
\Delta(x/z,\mu^2) \, ,
\eeqar
with $G$, $\Sigma$ and $\Delta$ being the twist-2 parton densities of
\Gl{sfunction}. The scale dependence of the coefficients $h_i$ is at variance
with the QCD predictions, as it is related to anomalous dimensions of twist-2
rather then twist-4 operators. In order to escape this difficulty we suggest
that \Gl{uvpart} should be
understood as a phenomenological model valid at some low scale $\mu^2$ of the
order of $1 - 2$ GeV$^2$, where higher twist contributions become important.
The mass scales $\La_{i,k}$ have to be fitted to experimental data.  The
non-singlet coefficients $A_{i,q}^{(NS)}$ were calculated in 
\cite{DMW96,SMMS,DW96,MMMSS}.
The pure-singlet coefficients $A_i^{(2)}$ (and
corresponding coefficients $A_i^{(4)}$, related to $1/Q^4$, or twist-6
corrections), which are the main result of the present paper, have been
collected in the appendix. 
 
\section{Calculation}

In this section we describe the calculation of the coefficients $A_i^{(2)}$
and $A_i^{(4)}$.  They can be obtained in many ways, either by calculating
one-loop Feynman diagrams with an infrared regulator such as a gluon mass
\cite{Ben95}, or using a dispersion representation of the 'effective' coupling
constant discussed in \cite{DMW96}, or by analysing the large-order behavior
of the twist-2 coefficient functions.  We follow the third method by
investigating the large-order behavior of the coefficient functions in the
limit that the number of flavours, i.e.~$n_f$, goes to infinity. In the case
of modeling power corrections to NS-structure functions it is sufficient to
compute one-loop diagrams, while the evaluation of the singlet part is more
complicated and involves two-loop calculation. The effective coupling method
has been recently used in the two-loop calculation of power corrections to
photon-photon scattering by Hautmann \cite{Hau98}. Up till now, the
equivalence of all three aforementioned methods of computing $A_i^{(2)}$ and
$A_i^{(4)}$ has been firmly established only at a one loop level. In case of a
two-loop calculation like the present one the equivalence between renormalon
approach and the finite gluon mass method persists, as discussed below, but
the relation to the effective coupling approach \cite{DMW96} has not yet been
proven.

The singlet coefficient may be decomposed as  
\beq{qsinglet}
C_{i,q}^S = C_{i,q}^{NS} + C_{i,q}^{PS} \quad,
\eeq
where $PS$ denotes the so-called 'pure-singlet' part where the photon line is
attached to a closed quark loop which is not sensitive to the flavour of the
target. While the large-$n_f$ expansion of $C_{i,q}^{NS}$ is already known,
here we describe the large-$n_f$ expansion of $C_{i,q}^{PS}$, see diagrams in
\Ab{fig1}.  To investigate the large-$n_f$ behavior of the gluonic coefficient
function $C_{i,G}$ one has to analyze fermion box diagrams with external gluon
legs where in addition a fermion bubble chain is inserted, see \Ab{fig2}.
\begin{figure}[tb]
\vspace{1cm}
\centerline{\psfig{figure=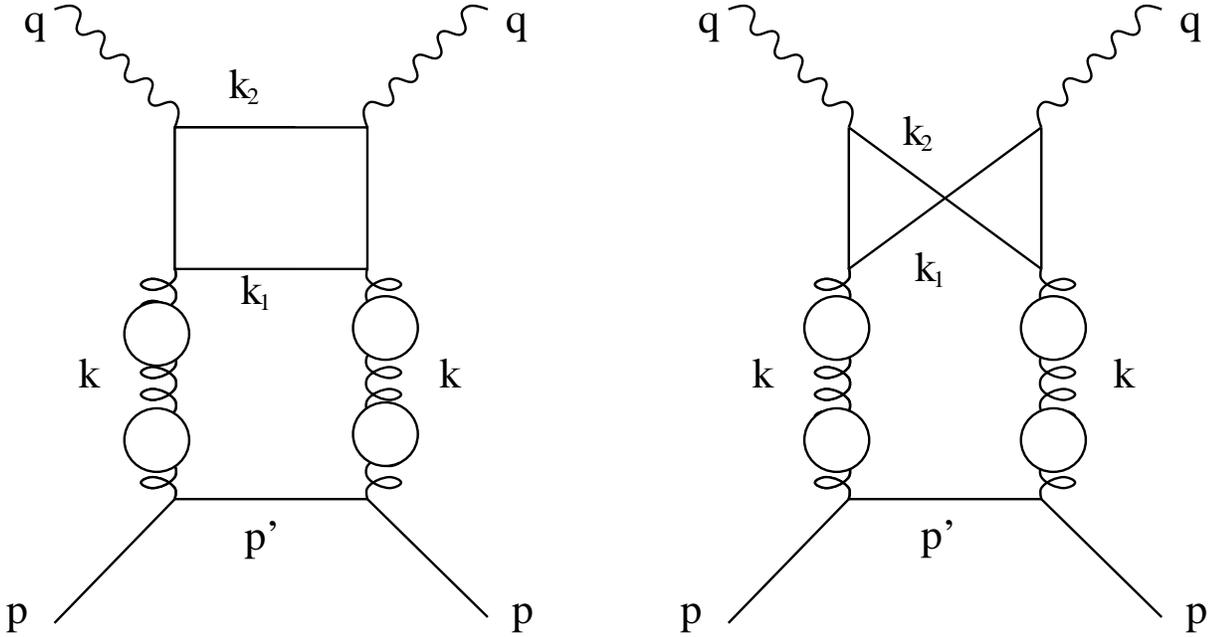,width=16cm}}
\caption[]{\sf Diagrams which contribute to the 'pure-singlet' part of 
the quark coefficient function.}
\label{fig1}
\end{figure}
\begin{figure}[tb]
\vspace{1cm}
\centerline{\psfig{figure=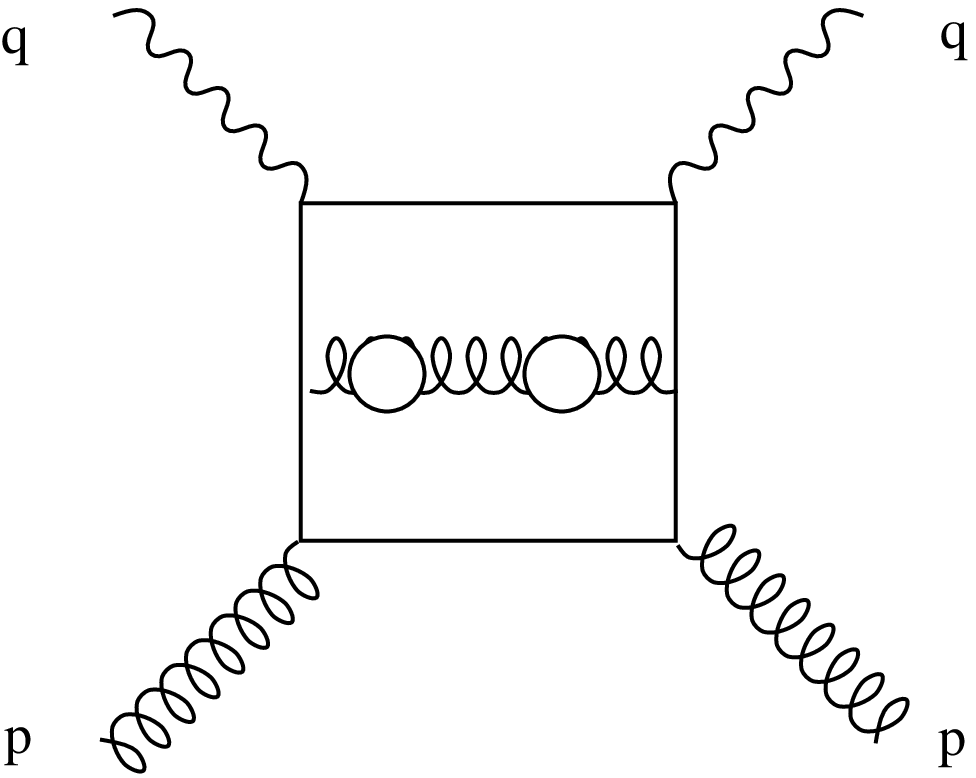,width=10cm}}
\caption[]{\sf Typical diagram which contributes to the   
gluon coefficient function.}
\label{fig2}
\end{figure}
Explicit enumeration of the $n_f$ factors arising in both cases shows that
contributions of the latter type are suppressed by a factor of $1/n_f$ as
compared to the quark pure-singlet part. The large-$n_f$ structure of the quark
pure-singlet coefficient can be written as
\beq{nfpures}
C_{i,q}^{PS} \sim
\left\{n_f a_s\right\} 
\bigg[\left(a_s n_f\right)^n a_s \bigg] \, .
\eeq
The curly brackets indicate the contribution in the $n_f$ expansion that comes
from the fermion box. The gluonic coefficient is formally of the same order in
$n_f$.  Taking into account that the lower vertex couples to the quark singlet
or gluon structure function one obtains effectively an additional factor of
$n_f$ in the former case.  The quark 
singlet structure function $\big(\frac{1}{n_f}\sum_{k=1}^{n_f} e_k^2 \big)
\Sigma(x,\mu^2)$ is ${\cal O} (n_f)$ as compared with the gluon
structure function $G(x,\mu^2)$\footnote{We thank V.Braun for useful
  discussions about this point.}. Recall, for example, that in the large-$n_f$
limit the fraction of longitudinal momentum carried by quarks and gluons is of
the order $1 - {\cal O}(1/n_f)$ and ${\cal O}(1/n_f)$, respectively 
\cite{Pes95}. Following this argument one has to 
calculate only the quark singlet part as the dominating contribution in the
large-$n_f$ limit.

Instead of calculating the $a_s$
corrections term by term to obtain the large-order behavior of the
perturbative expansion,  
it is convenient to deal directly with the Borel transform of the whole series.
In the special case we are considering here, there are two bubble
chains which contribute, both having the same virtuality 
$-k^2$, see \Ab{fig1}. By performing the substitution
$a_s^{n+1} \to u^n/n!$
one easily obtains the Borel transform of the square of the running coupling 
\beq{dborel}
BT\left[\left(\frac{a_s}{1 + \beta_0 a_s 
\log(-k^2/(\mu^2 e^{-C}))}\right)^2\right]
= \frac{s}{\beta_0} \left(\frac{\mu^2 e^{-C}}{-k^2}\right)^s
\; ,
\eeq
with $s = \beta_0 u$, and $u$ being the Borel parameter.
The Borel transform can be viewed as generating
function for fixed order coefficients. Since our expansion
starts at order $a_s^2$ the Borel transform vanishes at $s=0$ as 
it should be. 
Inserting \Gl{dborel} in the loop calculation this simple picture 
can be destroyed when collinear poles manifest itself as $1/s$ 
contributions which effectively simulate finite contributions to the 
Borel transform.
Those poles are first to be subtracted to allow for the determination 
of fixed order results.

At this point it is convenient to discuss the equivalence between the
renormalon approach and the finite gluon-mass method, where functions
$A_i^{(2)}$ and $A_i^{(4)}$ are identified as coefficients of non-analytic
terms in the small gluon-mass expansion of diagrams in \Ab{fig1}. Note that in
the present case both gluon chains depend on the same virtuality $k^2$. 
Following \cite{Zak94}, one can derive the Mellin representation of the
square of the massive gluon propagator in the form: 
\beq{mg} 
\frac{1}{(k^2-\lambda^2)^2} =
-\frac{1}{2 \pi i}\, \frac{1}{k^4}
\int\limits_{-1/2 -i \infty}^{-1/2 +i \infty} \dd u
\frac{\pi (u+1)}{\sin{(\pi u)}} \left(\frac{\lambda^2}{-k^2}\right)^u 
\eeq 
which can be rewritten as
\begin{equation} 
\frac{1}{(k^2-\lambda^2)^2} =
-\frac{\beta_0}{2 \pi i}\, 
\frac{1}{k^4}
\int\limits_{-1/2 -i \infty}^{-1/2 +i \infty} \dd u
\frac{\pi (u+1)}{u \sin{(\pi u)}} 
\left(\frac{\lambda^2}{\mu^2 e^{-C}}\right)^u 
BT[a_s(k^2)],
\end{equation} 
see \Gl{dborel}. Inserting this equation into \Gl{partensor} below and
inverting the resulting formula by means of the Laplace transformation, it is
easy to see that the one-to-one correspondence between renormalon poles and
non-analytic terms in the small gluon-mass expansion, discussed in
\cite{Ball95}, holds at the two-loop level as well. On the other hand,
relation between the renormalon method and the effective coupling approach still
requires further investigation \cite{InPreparation}.

To obtain the IR-renormalon poles from the diagrams in \Ab{fig1} 
we  found it advantageous to follow the procedure 
for calculating DIS coefficient functions explained 
in \cite{Zijl92,Alt79}. In this method the hadronic scattering
tensor is calculated directly as the cross section of photon-parton
scattering.
Recall that in the partonic picture the hadronic structure functions
can be written as 
\beq{sfcal} F_i^{tw-2}(x,Q^2) = x \sum_k \int_x^1 \frac{dz}{z}
\hat{\f}_k(x/z) \hat{\cal F}_{i,k}(z,Q^2,\ep) \quad,
\eeq 
where $\hat{\f}_k$ and $\hat{\cal F}_{i,k}$ denote the bare parton density and
the parton structure function corresponding to a parton $k = q,\bar{q},G$.
The parton structure function can be obtained order by order in the
perturbation 
theory by calculating the radiative corrections to the parton subprocess
\beq{sub} 
\gamma^*(q) + k(p)  \to X  \quad,
\eeq
where a virtual photon of momentum $q$ hits a parton $k$ 
with momentum $p$.  In the general case, $\hat{\cal F}_{i,k}$ 
contains collinear divergences.  They can be
regularized e.g., by performing the calculation in $d=4-2\ep$ dimensions.  The
dependence of $\hat{\cal F}_{i,k}$ on the regulator is indicated by the
parameter $\ep$.  The infrared-divergent part has to be subtracted from the
partonic cross-section and factored into the bare parton density to define a
physical, measurable quantity.  In the present calculation it is the Borel
parameter $s$ which alters the power of the gluon propagator, and therefore
regularizes divergent integrals. Because of that, it is sufficient to perform
the calculation in $d=4$ dimensions. The collinear $1/\ep$ poles then 
show up as $1/s$ poles. Since we are only interested in the power 
corrections which manifest themselves as poles at $s=1,2,\dots$, 
there is no need to calculate the subtractions explicitly. 
Note that the anomalous dimensions in the 
$\overline{MS}$ scheme do not exhibit renormalon
singularities, and therefore the subtraction terms do not introduce
additional renormalon poles \cite{Ball95,MME97}. 
Hence, the IR-poles in the Borel image of
structure functions $F_i(x,Q^2)$ arise only from the 
IR renormalon poles in the Borel image of partonic structure functions
${\cal F}_{i,k}$, calculated without subtractions.

Partonic structure functions can be obtained 
from the partonic tensor 
\beq{partensor}
\hat{W}_{\mu\nu}(p,q) = \frac{1}{2} \, \frac{1}{4 \pi} \sum_{l = 1}^\infty
\int d P S^{(l)} M_\mu(l)M_\nu(l)^* \quad,
\eeq
where $\int d P S^{(l)}$ represents the $l$-body 
phase space integration. The factor $\frac{1}{2}$ comes from averaging
over quark spins in the unpolarized case.
The tensor $\hat{W}_{\mu\nu}(p,q)$ allows for an 
equivalent decomposition in terms of
$\hat{\cal F}_{2,k}$, $\hat{\cal F}_{L,k}$, $\hat{\g}_{1,k}$
as the hadronic scattering tensor in \Gl{wmunu}.
Choosing appropriate projections it holds:
\beqar{proj}
\hat{\cal F}_{L,k} &=& \frac{8 z^2}{Q^2} p_\mu p_\nu \hat{W}^{\mu\nu}
\nn \\
\hat{\cal F}_{2,k} &=& - \left(\hat{W}^\mu_\mu - \frac{12 z^2}{Q^2}
p_\mu p_\nu \hat{W}^{\mu\nu}\right)
\nn \\
\hat{\g}_{1,k} &=& \frac{i}{2} \frac{1}{q\cdot p}
\varepsilon_{\mu\nu\la\et} q^{\la} p^{\et} \hat{W}^{\mu\nu} \quad.
\eeqar
The normalization is chosen such that in the simple parton model we
get
\beqar{norm}
\hat{\cal F}_{L,q}^{(0)}(z,Q^2) =
\hat{\cal F}_{L,G}^{(0)}(z,Q^2) = 
\hat{\cal F}_{2,G}^{(0)}(z,Q^2) = 0 \, ,
\nn \\
\hat{\cal F}_{2,q}^{(0)}(z,Q^2) = \delta(1-z) \, ,
\hat{\cal F}_{L,q}^{(1)}(z,Q^2) = a_s C_F (4 z) \, ,
\eeqar

\subsection{Quark coefficient functions}

To compute the pure-singlet contribution one needs to 
calculate the diagrams in \Ab{fig1}. In the improved parton model
this corresponds to the calculation of the two-to-three-body subprocess  
\beq{reaction}
\ga^*(q) + p \to k_1 + k_2 + p' \quad,
\eeq
which requires the three body phase space integrals
\beqar{pspace}
\int \dd P S^{(3)} &=& 
\frac{1}{(2\pi)^5} \int \dd^4 k_1  \dd^4 k_2  \dd^4 p'  
\delta^{(+)}(k_1^2) \delta^{(+)}(k_2^2)\delta^{(+)}(p'^2)
\\ \nn
&&\times \delta(p + q - k_1 - k_2 - p')
\eeqar
that appear in \Gl{partensor}.
It turns out to be convenient to evaluate the three particle
phase space in the CMS system of the outgoing $q\bar q$ pair 
of the fermion box. Introducing variables 
as in \cite{Ma89} we find for the phase space integral
\beqar{pspaceII}
\int d P S^{(3)} &=& 
\frac{1}{(4 \pi)^4} Q^2 \frac{1-z}{z} \int_0^1 {\rm d}y  (1-y)\int_0^1
{\rm d}v \int_0^\pi {\rm d}\phi \int_0^\pi {\rm d}\theta \sin(\theta)
\quad.
\eeqar
Here $\theta$ and $\phi$ are the polar and azimuthal angles 
in the CMS system of $k_1$ and $k_2$ respectively, while
$v$ and $y$ are rescaled $u$ and $t$ channel invariants.
The exact definition of theses variables can be found in the
appendix of \cite{Ma89}. 
For our purpose the definition of $y$
\beq{y}
y = \frac{-k^2}{Q^2} \cdot z \quad,
\eeq 
is important, where $-k^2$ is the virtuality of the dressed gluon 
and $z$ is the argument of the partonic structure function 
$\hat{\cal F}_{i,k}$. 
It follows that the Borel transforms of dimensionless projections of the 
partonic  tensor \Gl{proj} can be written in  the form
\beq{btwmunu}
BT[\hat{\cal F}(z)](s) = \int_0^1 \frac{\dd y}{z} 
\left(\frac{z}{y}\right)^{2+s} 
\frac{s}{\be_0} \left(\frac{\mu^2 e^{-C}}{Q^2}\right)^s 
\hat{\cal F}(z,y) \quad,
\eeq
which shows that the dependence on the Borel parameter
$s$ factorises and is influenced only by the $y$ integration.
To find the renormalon poles it is therefore not necessary to evaluate
the whole expression exactly. Instead, we can expand
$\hat{\cal F}(z,y)$ in $y$. 
The integration over $y$ transforms
terms $\sim y^n$ into simple poles $\sim s/(n-1-s)$ and terms
$\sim y^n \log(y)$ into double poles $\sim s/(n-1-s)^2$. Note
that $\log(y)$ terms arise from the integration
over the polar angle $\theta$  where partons $k_1$ and $k_2$ 
become collinear.

The expansion of $\hat{\cal F}(z,y)$ in $y$ then 
leads to an expression of the form
\beq{expy}
\hat{\cal F}(z,y) = \sum_{n=0}^\infty \left(A_n(z) + B_n(z) \log(y)\right)
\left(\frac{y}{z}\right)^n \quad.
\eeq
The expansion coefficients $A_n(z)$ and $B_n(z)$ can be expressed as linear
combinations
$$
A_n(z),B_n(z) \sim \log(z) w_{1,n}(z) + w_{0,n}(z)
$$ 
with the coefficients $w_{1,n}(z)$ and
$w_{0,n}(z)$ being simple polynomials in $z$.
Inserting 
\Gl{expy} back into \Gl{btwmunu} we observe that although the form of the phase
space integral
suggests that the Borel transform $BT[\hat{\cal F}(z)](s)$ is simply
proportional to 
$z^s$, after performing the $y$ integration one actually obtains 
an expression of the form 
\beqar{expyII}
BT[\hat{\cal F}(z)](s) &=& \Big[ {\rm const.} +
\left(A_0(z)-B_0(z)\right)\frac{z^{s+1}}{1+s} - 
B_1(z)\frac{z^{s}}{s} \nn \\ 
&&+
\left(A_2(z)+B_2(z)\right)\frac{z^{s-1}}{1-s} - 
B_2(z) \frac{z^{s-1}}{(1-s)^2} 
\nn \\
&&+
\left(2 A_3(z)+B_3(z)\right)\frac{z^{s-2}}{2-s} - 
2 B_3(z) \frac{z^{s-2}}{(2-s)^2} +
{\cal O}\left(\frac{1}{3-s}\right)  
\Big]\frac{1}{\beta_0}\left(\frac{\mu^2 e^{-C}}{Q^2}\right)^s \, .
\nn \\
\eeqar
It is interesting to note that the coefficient of the 
$1/(1+s)$-pole in the sum over all diagrams vanishes, 
i.e.~that the coefficient function of a DIS
structure function does not contain any UV-renormalon poles.
This is in line with previous observations that
positions of UV-renormalon poles in NS-structure functions always 
depend on the Mellin moment  $N$ \cite{SMMS}, and that 
therefore UV-renormalon contributions are absent after the
inverse Mellin transformation is performed.

Note the appearance of double poles in \Gl{expyII}, which were 
not present in one-loop calculations of renormalon contributions to
DIS structure functions in the non-singlet sector \cite{SMMS,DW96,MMMSS}. 
By considering renormalisation group improved version of 
OPE it can be shown that renormalon singularity of the Borel
transform is in general expected to have the form \cite{Mue85}
\beq{singborel}
BT[\hat{\cal F}(z)](s) \sim 
\frac{const.}{(n-s)^{1+\gamma_0/\be_0}} \; .
\eeq
Here $\gamma_0$ corresponds to an eigenvalue of one-loop
anomalous dimension matrix of higher twist operators of 
dimension $n$ which contribute to OPE of the hadronic tensor
in \Gl{wmunu}. Calculation of  
anomalous dimensions
considerably simplifies in the large-$n_f$ limit, resulting in eigenvalues
which are either equal to zero or to a 
multiple of $\be_0$. The double pole found in \Gl{expyII} arises from operators
with $\gamma_0 = \be_0$.

Terms in \Gl{expyII} singular at $s=0$ are directly related to singularities
which occur in a calculation of $\hat{\cal F}(z)$ in $d = 4 - 2 \epsilon$
dimensions. The coefficient of the $1/s$ pole corresponds to the
quadratic 
collinear divergence and is directly related to the $1/\ep^2$ pole
in the usual dimensional regularization. 
In general, this coefficient can be written as a convolution of
the leading order splitting functions
\beq{b1}
B_1(z) \sim P^{(0)}_{qg} \otimes  P^{(0)}_{gq} \, .
\eeq
Thus, for the structure function $F_2$ one obtains
\beq{bqf2}
B_1(z) \sim  P^{(0)}_{qg} \otimes  P^{(0)}_{gq}
= C_F T_F n_f\frac{1}{3 x} 
\left[ 4 (1-x)^3 + 3 x (1-x) + 6 (1+x) \log(x)\right] \;,
\eeq
while for $g_1$ one finds
\beq{bqg1}
B_1(z) \sim \Delta P^{(0)}_{qg} \otimes  \Delta P^{(0)}_{gq}
= 32\, C_F T_F n_f \left[5(1-x) + 2 (1+x) \log(x)\right] \; .
\eeq
Note that for the longitudinal projections the fermion box is free
of collinear divergences and therefore the $F_L$ contribution does not contain
any $1/\ep^2$ pole.
 
The form of the constant term in the expansion can also be determined from
general considerations. 
It is given by a combination of next to leading order splitting function and a
convolution of   
leading order coefficient function and leading order splitting
function corresponding to single $1/\ep$ poles. However,
determination 
of the constant term in our approach requires the expansion to all
orders given in \Gl{expy}.

Recall that for $x$ close to 1 one expects the ratio of twist-2 to twist-4
contributions to behave like $\Lambda^2/(Q^2(1-x))$, or approximately like
$\Lambda^2/W^2$ \cite{Bro79}. The same trend at small $x$ would correspond to
an expansion in terms of $x \, \Lambda^2/Q^2$, see e.g. Ref. \cite{For97}. 
However, the exact
calculation which takes into account \Gl{expyII} shows that in the small $x$
region the expansion parameter, see 
\Gl{irrenormalon} below, turns  
out to be $\Lambda^2/Q^2$ instead of
$x \, \Lambda^2/Q^2$. As a consequence, in the renormalon model the ratio of
the 
twist-4 and twist-2 or the twist-6 and twist-4 contributions goes to a
constant limit as $x \to 0$. Note that in the renormalon model calculation of
power corrections to 
parton fragmentation functions in $e^+ \, e^-$ anihilation into hadrons the
expansion parameter turns out to be $\Lambda^2/(x^2 Q^2)$. Hence, the
expansion breaks down in the region $x < \Lambda/Q$ and has to be resummed 
\cite{BBM97}. No such resummation is required in the present case. 

To obtain the coefficients $A_{i,k}$ in \Gl{uvpart} one has to transform the 
Borel images back to the $a_s$ representation by using the inverse Borel
transformation 
\beq{sum}
\hat{\cal F}(z;a_s) = \frac{1}{\be_0} \int\limits_0^\infty 
ds\,e^{-s/(\be_0 a_s)}
BT[\hat{\cal F}(z;a_s)](s) \; .
\eeq
An unambigious reconstruction of the perturbative expansion is 
prevented by the poles on the  
integration contour and therefore the summed series acquires an
imaginary part which is directly related to the UV-divergent part
of the higher twist operators we are looking for.
Its magnitude can be estimated either directly or by taking the imaginary part
(divided by $\pi$) $\Im/\pi$  of the inverse Borel transform in \Gl{sum}
\cite{Bra95}.  This procedure leads, for the simple and quadratic
poles of the right-hand side of \Gl{expyII}, to
\beqar{irrenormalon}
\frac{\Im}{\pi} \frac{1}{\be_0} \int\limits_0^\infty ds\,e^{-s/(\be_0 a_s)}
\frac{z^{s-n}}{(n-s)}
\frac{1}{\beta_0}\left(\frac{\mu^2 e^{-C}}{Q^2}\right)^s 
&=& \pm \frac{1}{\beta_0^2}\left(\frac{\La^2 e^{-C}}{Q^2}\right)^n  
\nn \\
\frac{\Im}{\pi} \frac{1}{\be_0} \int\limits_0^\infty ds\,e^{-s/(\be_0 a_s)}
\frac{z^{s-n}}{(n-s)^2}
\frac{1}{\beta_0}\left(\frac{\mu^2 e^{-C}}{Q^2}\right)^s 
&=& \pm \frac{1}{\beta_0^2}\left(\frac{\La^2 e^{-C}}{Q^2}\right)^n  
\log(\La^2 e^{-C} z/Q^2) \quad.
\eeqar
The ambiguity in the sign of the IR-renormalon contributions is due to
the two possible contour deformations, namely 
above or below the pole at $s = n$.

\subsection{Gluon coefficient functions}

As it has been mentioned above, the gluonic contribution is
subleading in the large-$n_f$ limit. For finite $n_f$, however, gluons
certainly 
play an important role. To model their contribution we have adopted
a
procedure introduced in \cite{BBM97} in the analysis 
of power corrections in fragmentation processes in $e^+ e^-$ annihilation.
The idea is to represent the pure-singlet unpolarized quark
coefficient function $A_{i,q}^{PS} $ as a convolution of the
leading order quark-gluon unpolarized splitting function 
\beq{split}
P^{(0)}_{gq}(z) = C_F \frac{1 + (1-z)^2}{z} 
\eeq
with an effective gluonic unpolarized coefficient function $A_{i,G}(z)$. 
After taking Mellin moments of both sides of the equation
\beq{defglue}
A_{i,G}\otimes P^{(0)}_{gq} = A_{i,q}^{PS}\, ,
\eeq
$A_{i,G}$ can be obtained by performing
an inverse Mellin transformation of
\beq{defglueII}
\tilde{A}_{i,G}(N) = \frac{\tilde{A}_{i,q}^S(N)}{\tilde{P}^{(0)}_{gq}(N)} \; .
\eeq
The same procedure can be applied also to define the polarized gluon
coefficient function $A_{\Delta, G}(z)$, with the obvious modification that now
the polarized pure-singlet quark coefficient function and the polarized
quark-gluon splitting function
\beq{polsplit}
\Delta P^{(0)}_{gq}(z) = 4 C_F (2-z)
\eeq
enter \Gl{defglueII}. 
The obtained formulae are presented in the appendix. In \Ab{glupsf2} and
\Ab{gluonpsg1} we
have compared 
the ratios of quark and gluonic twist-4 corrections to 
$F_2(x,Q^2)$ and to $g_1(x,Q^2)$, respectively. It turns out that in both
cases the shape of the
gluon contribution which results from deconvolution procedure described
above is quite similar to the shape of the original quark contribution. It
is also easy to show that the 
ratio of quark and gluon contributions tends to a constant
if $x \to 0$. As the overall normalisation of each contribution is anyhow an
adjustable fit
parameter,  we have modified the ratios in \Ab{glupsf2} and
\Ab{gluonpsg1} by dividing in each case the quark
contribution by a factor 2.

\begin{figure}[tb]
\vspace{1cm}
\centerline{\psfig{figure=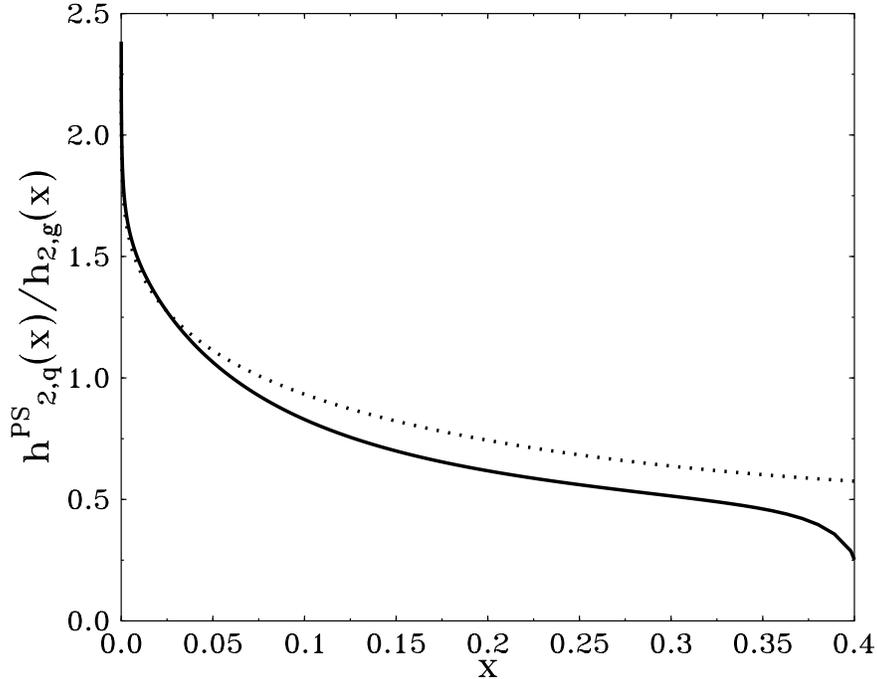,width=14cm}}
\caption[]{\sf 
Ratio of the quark pure-singlet contribution to the
gluonic contribution obtained according to \Gl{defglueII} for $F_2$.
The dotted line shows contributions of the 
single poles alone, the solid line corresponds to the
coherent sum of contributions from single and double poles. 
We have adjusted the normalization by dividing the pure-singlet
quark contribution by 2.}
\label{glupsf2}
\end{figure}

\begin{figure}[tb]
\vspace{1cm}
\centerline{\psfig{figure=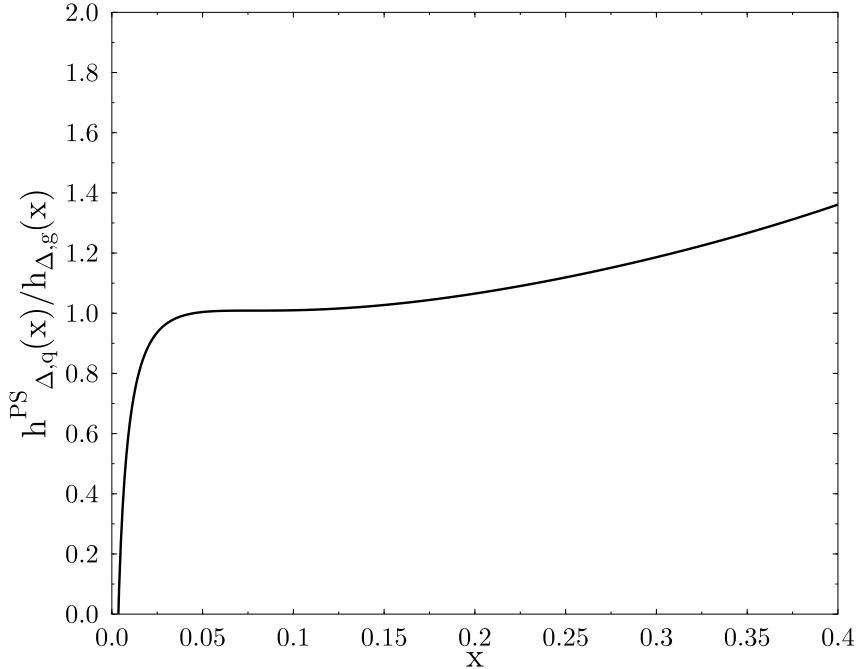,width=14cm}}
\caption[]{\sf 
Ratio of the quark pure-singlet contribution to the
gluonic contribution obtained according to \Gl{defglueII} for $g_1$. 
The solid line shows a coherent sum of contributions from single and double
poles.We have adjusted the normalization by dividing the pure-singlet
quark contribution by 2.}
\label{gluonpsg1}
\end{figure}

Note that this procedure results in the twist-4 contribution originated by
gluons which
depends linearly on the twist-2 gluon longitudinal momentum
distribution. On the other hand, it has been known for a long time that an
important 
class of twist-4 corrections to DIS originates from matrix elements of gluonic
operators involving four gluon  
field strength tensors \cite{Bar93}, thus being proportional to the square of
the gluon momentum distribution. 
This is a good illustration of the fact that the renormalon model does
not 
provide a description of all contributions to power suppressed corrections,
and that 
some of the missed terms can be important for phenomenology.

\section{Discussion}

As it is known that the renormalon model cannot account for the absolute
normalization of the twist-4 corrections, we follow the attitude proposed in
\cite{BBM97,MSSM}, and assume that model parameters,
the overall
normalization factors and mass scales, should be fitted to the data. 
Recall that in the flavour non-singlet case data at $x \ge 0.25$ can
be fitted using the normalization factor about 2 times larger than $\mu^2_R =
\frac{2}{\beta_0} \Lambda^2 e^{-C}$
arising directly from the renormalon model. To get a feeling about
the shape of pure flavour-singlet twist-4 corrections, in the following we
discuss 
the renormalon model predictions for twist-4 corrections to 
$F_2(x,Q^2)$, $g_1(x,Q^2)$, and $F_L(x,Q^2)$ for a 
deuteron target, obtained at a fixed mass scale. For definiteness, we have used
the NLO MRSA parameterization \cite{MRS95} and the 
NLO Gehrman-Stirling gluon-A set \cite{GS95} at $Q^2
= 4$ GeV$^2$ as an input
for twist-2 unpolarized and polarized parton distributions, respectively. We
have checked that other sets of LO or NLO parton distributions produce similar
results. While such an analysis cannot substitute for a data fit, it
transparently reveals
the gross features of the model. 

\begin{figure}[tb]
\vspace{1cm}
\centerline{\psfig{figure=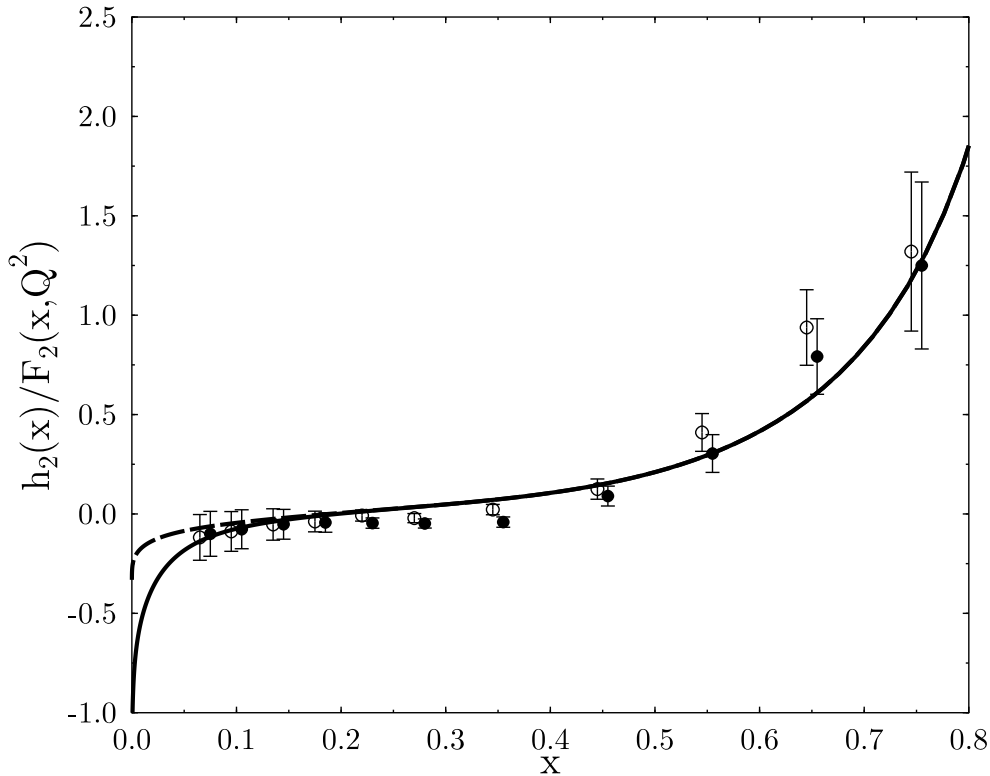,width=14cm}}
\caption[]{\sf Renormalon model prediction for the ratio 
$h_2(x)/F_2(x,Q^2)$ for deuteron and
proton targets.
The full line shows the non-singlet and pure-singlet quark
contributions. The dashed line corresponds to the non-singlet and gluonic
contributions. The filled and empty circles denote data points for
proton 
and deuteron, respectively \cite{VM92}. The magnitude of the non-singlet
contribution 
has been normalized to the large-$x$ data.
$\Lambda_{\overline{MS}}=200~{\rm MeV}$, $Q^2=4~{\rm GeV}^2$ and $n_f = 4$.
}
\label{fig:ratiof2}
\end{figure}

\begin{figure}[tb]
\vspace{1cm}
\centerline{\psfig{figure=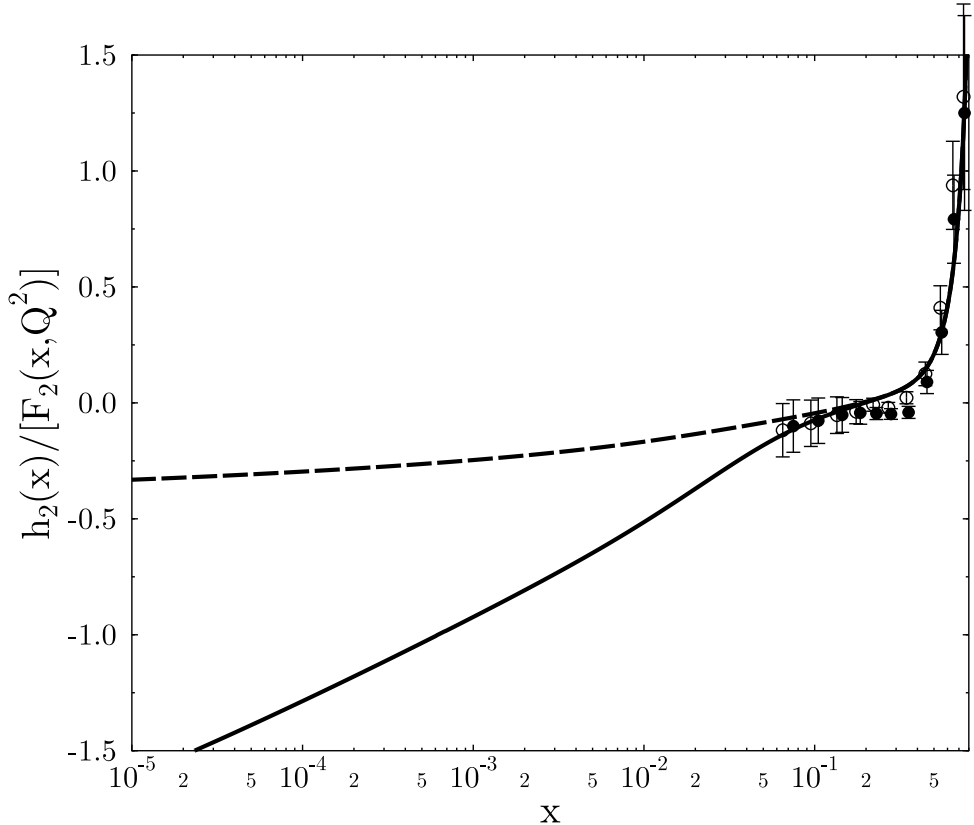,width=14cm}}
\caption[]{\sf 
The same as in \Ab{fig:ratiof2}, but with a logarithmic scale on
the $x$-axis
}
\label{fig:ratiof2_smallx}
\end{figure}

Fig.~\ref{fig:ratiof2} shows the prediction for the ratio
$F_2^{\rm tw-4}(x,Q^2)/F_2^{\rm tw-2}(x,Q^2)$ at $Q^2 = 4$ GeV$^2$ for a
deuteron target as compared with the data from Ref.~\cite{VM92}. To
obtain this 
plot we have fixed the sign in the right-hand side of
\Gl{irrenormalon}, such that contributions arising from both simple and
quadratic poles add coherently. At large $x$ the renormalon model prediction
is dominated by 
contributions from NS-type graphs, common for singlet and non-singlet
channels, see \Gl{qsinglet}. It explains why it has been possible to
fit deuteron and proton data in this domain by the renormalon
model correction based only on the non-singlet coefficient function
\cite{DW96}. On the other 
hand, pure-singlet  
contributions dominate for
$x$ smaller than 0.2. In the limit
$x \to 0$ the ratio of twist-4 and twist-2 contributions tends to be a
constant. We note that the contribution from the quadratic pole turns out to be
larger than the contribution from the single one. If the pure-singlet quark
contribution is replaced by a gluonic one, a very similar shape of twist-4
correction results, in accordance with the discussion in section 3.2.
Figure \ref{fig:ratiof2_smallx} shows the same ratio with a logarithmic 
scale of the $x$-axis.

\Ab{fig:flsinglet} shows $h_L(x)$, the renormalon correction to $F_L$ plotted
against   
the coefficient of the $1/Q^2$ dependent term in the data fit of 
\cite{Whi90}, for a deuteron target. 
The target mass correction taken from \cite{San91}, 
makes up to 15\%-20\% of $h_L(x)$ as compared to the phenomenological fit.
The pure-singlet part becomes considerable only in the
small $x$ region, where it produces a steep raise, clearly at variance with
the parameterization of Ref.~\cite{Whi90}.

\begin{figure}[tb]
\vspace{1cm}
\centerline{\psfig{figure=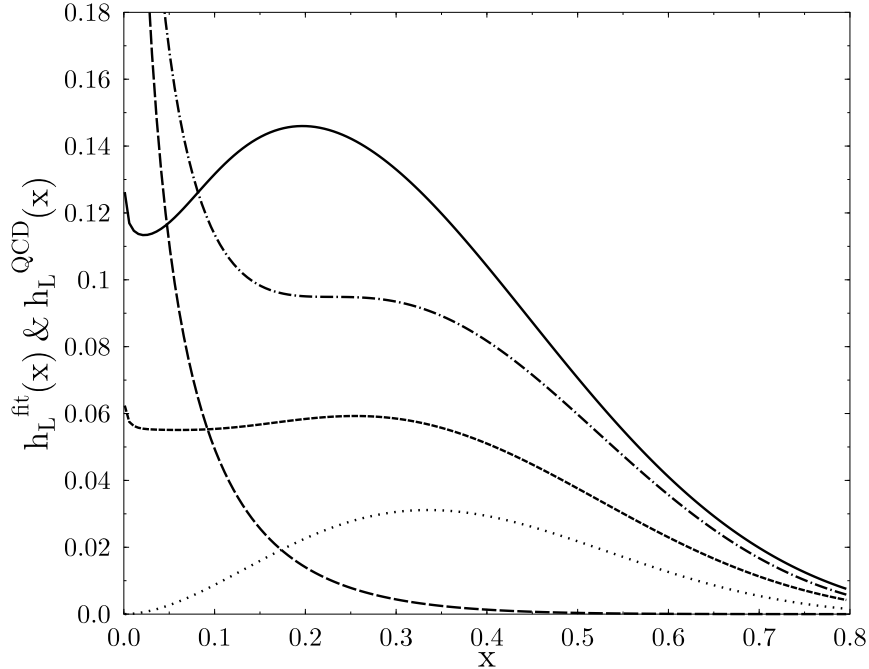,width=14cm}}
\caption[]{\sf 
The shape of the $1/Q^2$ power correction $h_L(x)$
to the longitudinal structure function $F_L(x,Q^2)$ for a deuteron target.
The long-dashed line depicts the pure-singlet quark
correction, with contributions from single and
double poles added coherently.
The short-dashed line shows the contribution of the non-singlet part.
The dotted line is the target mass
correction, taken from \cite{San91}. The dot-dashed line is the sum of all
contributions. 
The solid line represents $h_L(x)$ as obtained from 
the Whitlow-fit to $R(x)$ \cite{Whi90}
and the parameterization of $F^d_2$ from the 
NMC-analysis \cite{NMC95}.
For consistency, the renormalon model prediction shown here has been obtained
using the same parameterization of twist-2 parton densities.
$\Lambda_{\overline{MS}}=250~{\rm MeV}$, $Q^2=4~{\rm GeV}^2$ and $n_f = 4$.
}
\label{fig:flsinglet}
\end{figure}
\begin{figure}[tb]
\vspace{1cm}
\centerline{\psfig{figure=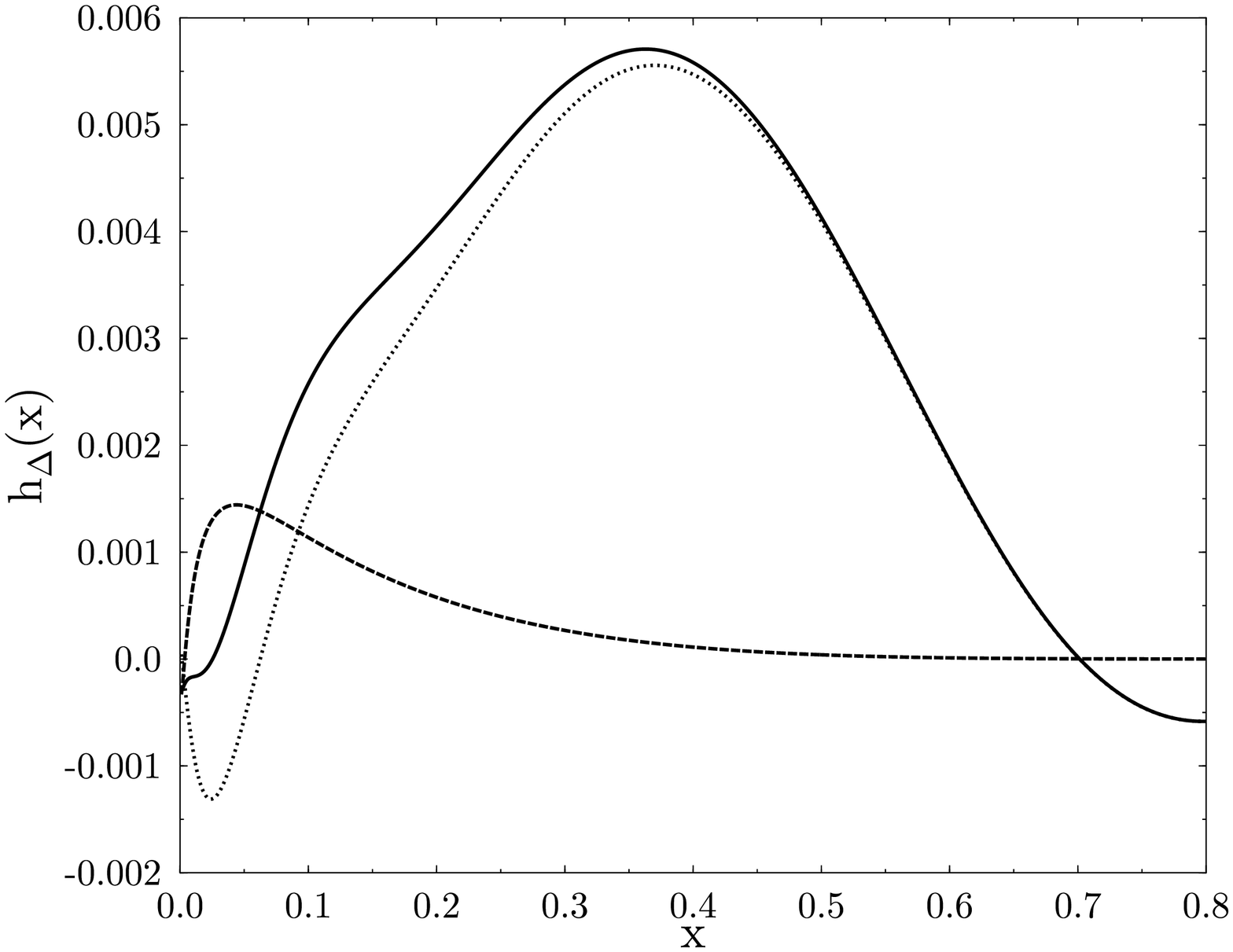,width=14cm}}
\caption[]{\sf 
The shape of the $1/Q^2$ power correction $h_\Delta(x)$
to the polarized structure function $x g_1(x,Q^2)$ for a deuteron target.
The dashed line corresponds to the  pure-singlet quark contribution, 
the dotted line to the contribution of the non-singlet part.
The full line is the sum of both.
$\Lambda_{\overline{MS}}=231~{\rm MeV}$, $Q^2=4~{\rm GeV}^2$ and $n_f = 4$.
}
\label{fig:g1singlet}
\end{figure}
As expected, in the unpolarized case the pure-singlet contribution is
negligible 
at large $x$, but it dominates when $x$ becomes small. Although, strictly
speaking, the absolute 
normalization is unknown, the renormalon model clearly suggests that while the
non-singlet 
twist-4 
contribution becomes smaller and smaller with $x$ decreasing, the magnitude of
the pure-singlet contribution 
raises for $x$ below 0.2, and therefore twist-4 corrections should not be
neglected in this region. 

Recall that at large $x$ the UV 
dominance hypothesis predicts, in accordance with arguments based on an
analysis of multiparton contribution of hadronic wave functions \cite{Bro79}, 
that power corrections to $F_2(x,Q^2)$ are effectively
suppressed by $1/(Q^2(1-x)) \sim 1/W^2$, and
therefore they can be eliminated by a suitable $W^2$ cut-off. In the small-$x$
domain the situation is different. Our calculation predicts that
the ratio of twist-4 to twist-2 terms in this region depends rather weakly on
$x$, such that the former cannot be minimized by a 
kinematical cut.

Here a 
natural 
question arises to which extent the renormalon model predictions 
can be trusted in
the small $x$ domain. Although it is formulated in the field-theoretical
framework, the model 
almost certainly misses some important QCD physics. 
We note e.g. that ladder corrections which are known to
influence strongly the small $x$ behavior of twist-2 structure
functions are
absent here, as they are formally suppressed by $1/n_f$. Hence, while we
expect that general trends followed by twist-4 contributions are
correctly reproduced, it is
possible that details of small $x$ behavior of twist-4 corrections are not
properly described in this approach. 

Fig.~\ref{fig:g1singlet} shows the renormalon model prediction for the twist-4
part 
of the polarized structure function
$xg_1^{\rm tw-4}(x,Q^2)$. The resulting NS correction is negative at
small $x$, changes sign
around $x = 0.02$, and reaches a maximum around $x\sim
0.4$. The pure-singlet contribution is significant only in the small $x$
region but, contrary to the unpolarized case, even there it does not dominate
over the NS 
part.  As in the $F_2$ case the ratio of twist-4 and twist-2 contributions
to $g_1(x,Q^2)$ tends to be a constant for $x \to 0$.
In order to get a maximal magnitude of the pure-singlet correction, in
Fig.~\ref{fig:g1singlet} we have  
counted contributions from single and quadratic poles coherently. 

\section{Summary}

The calculation presented in this paper
extends previous renormalon model analyses of twist-4 corrections to DIS
\cite{DMW96,SMMS,DW96,MMMSS} to the 
flavour singlet sector. We have presented a sample of predictions for the
$x$-dependence of twist-4 corrections to nucleon structure functions
$F_2(x,Q^2)$, $g_1(x,Q^2)$, and $F_L(x,Q^2)$ in the whole $x$ domain. 
Our results 
suggest that in the small-$x$ region the twist expansion parameter is not $x
\cdot \Lambda^2/Q^2$, as it would follow from a $1/W^2$-dependence of higher
twist corrections, but is approximately equal to
$\Lambda^2/Q^2$.  In this case higher twists cannot be eliminated by a $W^2$
cut, and should be taken into account in the interpretation of the current
data. 

One
possibility is to fit the data taking into account, in addition to
twist-2 
contribution,
twist-4 corrections either in the form derived in the present paper with the
non-singlet contribution taken from \cite{DW96,MMMSS}, or as given in
Ref.~\cite{MarRys98}.

To obtain the renormalon model predictions for twist-4 corrections to nucleon
structure functions in the flavour-singlet sector one should insert the
pure-singlet coefficient functions, listed in the 
appendix below, into the right-hand side of \Gl{uvpart}, and add the
non-singlet type contribution. For each
contribution, the overall factor and its mass scale $\Lambda^2$ should be
fitted independently. In particular, mass scales can be different for different
channels. In principle one should also try a fit in which overall signs 
differ between contributions arising from single and quadratic poles. Because,
as discussed in section 3.2, shapes of the resulting quark and gluon
contributions are rather 
similar, such a fit could involve, besides standard
parameterizations 
of the twist-2 structure functions at some low virtuality, only two 
additional elements: the quark non-singlet term, which can be taken
from \cite{DW96}, and the gluonic term, as given below. Description of
twist-4 corrections requires then the introduction of 3 new parameters:
the normalization of the non-singlet quark
contribution and the normalization and the mass scale which enters the gluonic
contribution. In the currently most interesting case of $F_2(x,Q^2)$ a further
simplification is possible: 
as the flavour non-singlet part is important only in the large-$x$ domain, it
can be 
neglected in a fit to the small-$x$ data, which would reduce the number of
additional fit parameters to 2.

\vspace*{5mm}

{\bf Acknowledgements.} This work has been  
supported by BMBF and DFG. 
We are indebted to V. Braun, L. Magnea, W. van Neerven and S. Forte 
for numerous interesting discussions.
E.S. thanks DFG for support and the Department of Theoretical 
Physics of the University of Turin for the kind hospitality.

\section{Appendix} 

In this section we quote explicitly our renormalon model results for the
pure-singlet coefficient functions 
$A_{i,q}^{PS\;(2)}(z,\log(\La^2/Q^2))$ and $A_{i,G}^{(2)}(z,\log(\La^2/Q^2))$,
which control quark and gluon contributions to ${\cal O}(1/Q^2)$ corrections
to nucleon structure  
functions in the flavour singlet sector, see \Gl{uvpart}. For completeness, we
quote also the coefficient functions $A_{i,q}^{PS\;(4)}(z,\log(\La^2/Q^2))$ 
and 
$A_{i,G}^{(4)}(z,\log(\La^2/Q^2))$, arising from
the second IR renormalon at $s = 2$, which can be used to model 
${\cal O}(1/Q^4)$ corrections.

To facilitate comparison with the data which treats contributions from single
and quadratic  
poles separately, we have splitted each coefficient function into
contributions
arising from the single and quadratic poles, respectively. 

\subsection{Quark pure-singlet coefficient functions}
Coefficient function for the quark pure-singlet renormalon contribution to
$F_2$: The contribution of the quadratic pole is proportional to 
$\log(\La^2 z/Q^2)$.
\beqar{psf2tw4}
A_{2,q}^{PS\;(2)}(z,\log(\La^2/Q^2)) &=&
C_F\frac{T_F n_f}{\be_0^2}\frac{1}{z}
\Bigg[\bigg(
( \frac{32}{5} \log(z) - \frac{592}{75})z^5 
- (\frac{200}{9}-\frac{40}{3}\log(z)) z^3 
\nn \\ &&
+ (\frac{88}{3} + 24 \log(z)) z^2
+ (\frac{176}{225} - \frac{64}{15} \log(z)) \bigg)\Bigg]
\nn \\ 
&& +
C_F\frac{T_F n_f}{\be_0^2}\frac{1}{z}\Bigg[ 
\log(\La^2 z/Q^2)\bigg(
- \frac{32}{5} z^5 
+(48 \log(z) - \frac{184}{3}) z^3 
\nn \\ && 
+ (72 + 16 \log(z)) z^2 
- \frac{64}{15}\bigg)\Bigg]
\nn \\
\\
A_{2,q}^{PS\;(4)}(z,\log(\La^2/Q^2)) &=&
C_F\frac{T_F n_f}{\be_0^2} \frac{1}{z}
\Bigg[\bigg(
( - {\displaystyle \frac {128}{35}} \,
{\rm log}(z) + {\displaystyle \frac {3616}{3675}} )\,z^{7} + (
{\displaystyle \frac {192}{5}} \,{\rm log}(z) - {\displaystyle 
\frac {5792}{75}} )\,z^{5} 
\nn \\ &&
+ (64\,\log(z) + {\displaystyle 
\frac {128}{3}} )\,z^{4}+ ({\displaystyle \frac {64}{3}} \,\log(z) + 
{\displaystyle \frac {256}{9}} )\,z^{3} 
\nn \\ && 
+ ( - {\displaystyle 
\frac {64}{5}} \,\log(z) + {\displaystyle \frac {416}{75}} )
\,z^{2}  
+{\displaystyle \frac {512}{105}} \,\log(z) - 
{\displaystyle \frac {4576}{11025}} \bigg)\Bigg]
\nn \\
&& + 
C_F\frac{T_F n_f}{\be_0^2} \frac{1}{z}\Bigg[
64\log(\La^2 z/Q^2)\bigg(
z^{4} - {\displaystyle \frac {1}{3}} \,z^{3} - {\displaystyle 
\frac {1}{5}} \,z^{2} + {\displaystyle \frac {8}{105}}  - 
{\displaystyle \frac {3}{5}} \,z^{5} + {\displaystyle \frac {2}{
35}} \,z^{7}
\bigg)\Bigg]
\nn \\ 
\eeqar
Coefficient function for the quark pure-singlet renormalon contribution to
$F_L$:  
The first two lines contain the contribution from the single pole, 
the last line contains the contribution of the
quadratic pole:
\beqar{afl}
A_{L,q}^{PS\;(2)}(z,\log(\La^2/Q^2))& =&
C_F \frac{T_F n_f}{\be_0^2}\frac{1}{z}
\Bigg[\frac{32}{225}\bigg(
( 45 \log(z) - 33)z^5 
\nn \\ &&+ 
 (125-75\log(z)) z^3 - 75 z^2 - 30 \log(z) -17
\bigg)\Bigg]
\nn \\ 
&&+
C_F \frac{T_F n_f}{\be_0^2}\frac{1}{z}\Bigg[
\frac{32}{15}\log(\La^2 z/Q^2)
\bigg( - 3 z^5 - (10-15\log(z)) z^3 + 15 z^2 - 2\bigg)\Bigg]
\nn \\
\\
A_{L,q}^{PS\;(4)}(z,\log(\La^2/Q^2)) &=&
C_F\frac{T_F n_f}{\be_0^2}\frac{1}{z}
\Bigg[\frac{128}{15}\bigg(
( \frac{43}{735} - \frac{4}{7} \log(z))z^7 
 + (4\log(z)-\frac{94}{15})z^5 
\nn \\ && + (\frac{10}{3} + 5 \log(z)) z^4 + 5 z^3 - (\frac{32}{15} +
2 \log(z))z^2
+ \frac{3}{7} \log(z) + \frac{2}{245}
\bigg)\Bigg]
\nn \\ \nn \\
&& + C_F\frac{T_F n_f}{\be_0^2}\frac{1}{z}\Bigg[
\frac{128}{105}\log(\La^2 z/Q^2)
\bigg( 4 z^7 - 28 z^5 + 35 z^4 - 14 z^2 + 3 \bigg)\Bigg] \nn \\
\eeqar
Coefficient function for the quark pure-singlet renormalon contribution to
$g_1$:
The contribution of the 
quadratic pole is proportional to $\log(\La^2 z/Q^2)$.
\beqar{ag1}
A_{\Delta,q}^{PS\;(2)}(z,\log(\La^2/Q^2)) &=&
C_F\frac{T_F n_f}{\be_0^2}
\Bigg[\bigg(
 (\frac{80}{9}-\frac{16}{3}\log(z))z^3
+(12-12\log(z))z^2
\nn \\ &&
\qquad  \qquad \qquad
+(-12-12\log(z))z-\frac{16}{3}\log(z)-\frac{80}{9}
\bigg)\Bigg]
\nn \\ 
&& + 
C_F\frac{T_F n_f}{\be_0^2}\Bigg[
\log(\La^2 z/Q^2)\bigg(
\frac{16}{3}z^3+
(-8\log(z)+4)z^2
\nn \\ &&
\qquad \qquad \qquad
+(-4-8\log(z))z-\frac{16}{3}
\bigg)\Bigg] \nn \\  \\
A_{\Delta,q}^{PS\;(4)}(z,\log(\La^2/Q^2)) &=&
C_F\frac{T_F n_f}{\be_0^2} 
\Bigg[\bigg(
 (-\frac{272}{225}+\frac{32}{15}\log(z))z^5
+(\frac{208}{9}-\frac{32}{3}\log(z))z^3
\nn \\ &&
\qquad \qquad \qquad
+(-\frac{32}{3}\log(z)-\frac{208}{9})z^2
+\frac{272}{225}+\frac{32}{15}\log(z)
\bigg)\Bigg]
\nn \\ 
&& + 
C_F\frac{T_F n_f}{\be_0^2} \Bigg[
16 \log(\La^2 z/Q^2)\bigg(
-\frac{32}{15}z^5
+\frac{32}{3}z^3-\frac{32}{3}z^2+\frac{32}{15}
\bigg)\Bigg]
\nn \\ \eeqar
\subsection{Gluon coefficients functions}
In order to distinguish between contributions arising from 
single and quadratic poles we have split the coefficient functions
explicitly into two
parts, both proportional to $\frac{T_F n_f }{\be_0^2}$.
In addition, to keep the resulting expressions as compact as possible, we have
introduced the notation
\beqar{sundcfunction}
{\rm c}(z) &=& 
\sqrt{z}\,{\rm cos}\left(
{\displaystyle \frac {1}{2}} \,\sqrt{7}\,{\rm log}(z)\right)
\nn \\
{\rm s}(z) &=& \sqrt{\frac{z}{7}}\,{\rm sin}
\left({\displaystyle \frac {1}{2}} \,\sqrt{7}\,{\rm log}(z)\right) \quad,
\eeqar
which appear often in the process of
deconvoluting the unpolarized quark-gluon splitting function
from the pure-singlet coefficient function.

Gluon coefficient functions for $F_2$:
\beqar{gluef2t4}
A_{2,G}^{(2)}(z,\log(\La^2/Q^2)) &=&\frac{T_F n_f}{\be_0^2}\Bigg[
  {\frac{-55\,{\rm c}(z)}{7}} - {\frac{253\,{\rm s}(z)}{7}} +
   {\frac{32}{15\,z}} - 24\,z 
\nn \\ && +
   {z^4}\,\left( {\frac{912}{35}} - {\frac{192\,\log (z)}{7}} \right)
+
   {z^2}\,\left( {\frac{35}{3}} - 20\,\log (z) \right)\Bigg]
\nn \\
&+&
\frac{T_F n_f }{\be_0^2}\Bigg[
  {\frac{-102\,{\rm c}(z)}{7}} - {\frac{706\,{\rm s}(z)}{7}} +
   {\frac{32}{15\,z}}
\nn \\ && +
z\,\left( -104 - 32\,\log (z) \right)  +
   {z^4}\,\left( {\frac{272}{35}} + {\frac{192\,\log (z)}{7}} \right)
\nn \\ && +
   {z^2}\,\left( {\frac{326}{3}} - 64\,\log (z) - 72\,{{\log (z)}^2} \right)
\nn \\ && +
\log (\Lambda^2/Q^2)\,
    \left( {\frac{-122\,{\rm c}(z)}{7}} + 46\,{\rm s}(z) - 16\,z +
      {\frac{192\,{z^4}}{7}} + \right.
\nn \\ &&  \qquad \qquad \qquad  + \left.
{z^2}\,\left( 14 - 72\,\log (z) \right)
      \right) \Bigg]
\nn \\
\eeqar
\beqar{gluef2t6}
A_{2,G}^{(4)}(z,\log(\La^2/Q^2)) &=&\frac{T_F n_f}{\be_0^2}\Bigg[
  {\frac{31\,{\rm c}(z)}{28}} + {\frac{1621\,{\rm s}(z)}{28}} - 
   {\frac{256}{105\,z}} + {\frac{64\,z}{5}} + 
   {z^3}\,\left( -216 - 192\,\log (z) \right)  
\nn \\ && +
   {z^4}\,\left( {\frac{9952}{35}} - {\frac{1152\,\log (z)}{7}} \right)  + 
   {z^2}\,\left( -{\frac{232}{3}} - 32\,\log (z) \right)  
\nn \\ && +
   {z^6}\,\left( -{\frac{347}{140}} + 24\,\log (z) \right) \Bigg]
\nn \\
&+& 
\frac{T_F n_f}{\be_0^2}\Bigg[
  {\frac{9\,{\rm c}(z)}{28}} + {\frac{2019\,{\rm s}(z)}{28}} - 
   {\frac{256}{105\,z}} + {\frac{64\,z}{5}} 
\nn \\ && +
   \left( {\frac{136\,{\rm c}(z)}{7}} - 24\,{\rm s}(z) + 
      32\,{z^2} - 192\,{z^3} + {\frac{1152\,{z^4}}{7}} - 24\,{z^6} \right) \,
    \log (\La^2/Q^2) 
\nn \\ && +
   {z^3}\,\left( -88 - 192\,\log (z) \right)  + 
   {z^6}\,\left( -{\frac{557}{140}} - 24\,\log (z) \right)  + 
   {z^2}\,\left( {\frac{104}{3}} + 32\,\log (z) \right)  
\nn \\ && +
   {z^4}\,\left( {\frac{1632}{35}} + {\frac{1152\,\log (z)}{7}} \right) 
    \Bigg]
\nn \\
\eeqar
Gluon coefficient functions for $F_L$:
\beqar{glueflt4}
A_{L,G}^{(2)}(z,\log(\La^2/Q^2)) &=&\frac{T_F n_f}{\be_0^2}\Bigg[
  {\frac{-36\,{\rm c}(z)}{7}} - {\frac{204\,{\rm s}(z)}{7}} + 
   {\frac{32}{15\,z}} + {z^4}\,\left( {\frac{432}{35}} - 
      {\frac{192\,\log (z)}{7}} \right)  
\nn \\ && +
   {z^2}\,\left( -{\frac{28}{3}} + 16\,\log (z) \right) \Bigg]
\nn \\
&& +
\frac{T_F n_f}{\be_0^2}\Bigg[
  {\frac{-130\,{\rm c}(z)}{7}} - {\frac{342\,{\rm s}(z)}{7}} + 
   {\frac{32}{15\,z}} - 32\,z 
\nn \\ && + 
   \log (\La^2/Q^2)\,
    \left( {\frac{-52\,c(z)}{7}} + 28\,{\rm s}(z) + 
      {\frac{192\,{z^4}}{7}} + {z^2}\,\left( -20 - 48\,\log (z) \right) 
       \right)  
\nn \\ && + 
{z^4}\,\left( {\frac{272}{35}} + 
      {\frac{192\,\log (z)}{7}} \right)  + 
   {z^2}\,\left( {\frac{122}{3}} - 72\,\log (z) - 48\,{{\log (z)}^2} \right) 
\Bigg]
\nn \\
\eeqar
\beqar{glueflt6}
A_{L,G}^{(4)}(z,\log(\La^2/Q^2)) &=&\frac{T_F n_f}{\be_0^2}\Bigg[
  3\,{\rm c}(z) + {\frac{263\,{\rm s}(z)}{7}} - 
   {\frac{64}{35\,z}} + {\frac{256\,z}{15}} - 64\,{z^2} 
\nn \\ && + 
   {z^4}\,\left( {\frac{2816}{15}} - {\frac{1024\,\log (z)}{7}} \right)  + 
   {z^3}\,\left( -144 - 128\,\log (z) \right)  
\nn \\ && \qquad\qquad\qquad
+    {z^6}\,\left( {\frac{71}{35}} + 32\,\log (z) \right) 
\nn \\
&+&
\frac{T_F n_f}{\be_0^2}\Bigg[
  {\frac{51\,{\rm c}(z)}{7}} + {\frac{465\,{\rm s}(z)}{7}} - 
   {\frac{64}{35\,z}} + {\frac{256\,z}{15}} 
\nn \\ && + 
   \left( {\frac{96\,{\rm c}(z)}{7}} - 32\,{\rm s}(z) - 
      128\,{z^3} + {\frac{1024\,{z^4}}{7}} - 32\,{z^6} \right) \,
    \log (\La^2/Q^2) 
\nn \\ && + 
   {z^3}\,\left( -{\frac{176}{3}} - 128\,\log (z) \right)  + 
   {z^6}\,\left( -{\frac{557}{105}} - 32\,\log (z) \right)  
\nn \\ && + 
   {z^4}\,\left( {\frac{4352}{105}} + {\frac{1024\,\log (z)}{7}} \right) 
\Bigg]
\nn \\
\eeqar
Gluon coefficient functions for $g_1$:
\beqar{glueg1t4}
A_{\Delta,G}^{(2)}(z,\log(\La^2/Q^2)) &=&\frac{T_F n_f}{\be_0^2}\Bigg[
  {\frac{2}{3}} - 3\,z + {z^3}\,
    \left( -{\frac{44}{3}} + 8\,\log (z) \right)  
\nn \\ && + 
   {z^2}\,\left( 16 + 3\,\log (z) + 3\,{{\log (z)}^2} \right) \Bigg]
\nn \\
&& +
\frac{T_F n_f}{\be_0^2}\Bigg[
  {\frac{2}{3}} + {z^3}\,\left( {\frac{4}{3}} - 8\,\log (z) \right)  + 
   z\,\left( -9 - 4\,\log (z) \right)  
\nn \\ && +
   {z^2}\,\left( 7 + \log (z) + 5\,{{\log (z)}^2} + 
      {\frac{4\,{{\log (z)}^3}}{3}} \right)  
\nn \\ && +
   \log (\La^2/Q^2)\,
    \left( -2\, z - 8\,{z^3} + 
      z^2\,\left( 9 + 4\,\log (z) + 2\,{{\log (z)}^2} \right) 
       \right) 
\Bigg]
\nn \\
\eeqar
\beqar{glueg1t6}
A_{\Delta,G}^{(4)}(z,\log(\La^2/Q^2)) &=&\frac{T_F n_f}{\be_0^2}\Bigg[
  -{\frac{4}{15}} + {z^5}\,\left( {\frac{8}{5}} - 
      {\frac{32\,\log (z)}{9}} \right)  
\nn \\ && +
   {z^3}\,\left( -{\frac{112}{3}} + 16\,\log (z) \right)  + 
   {z^2}\,\left( 36 + {\frac{176\,\log (z)}{9}} + 
      {\frac{8\,{{\log (z)}^2}}{3}} \right) \Bigg]
\nn \\
&+& 
\frac{T_F n_f}{\be_0^2}\Bigg[
  -{\frac{4}{15}} + {z^3}\,\left( {\frac{8}{3}} - 16\,\log (z) \right)  + 
   {z^5}\,\left( {\frac{56}{135}} + {\frac{32\,\log (z)}{9}} \right)  
\nn \\ && +
   {z^2}\,\left( -{\frac{76}{27}} + 8\,\log (z) + 
      {\frac{8\,{{\log (z)}^2}}{3}} \right)  
\nn \\ && +
   \log (\La^2/Q^2)\,
    \left( -16\,{{\Mfunction{z}}^3} + {\frac{32\,{z^5}}{9}} + 
      {{\Mfunction{z}}^2}\,\left( {\frac{112}{9}} + 
         {\frac{16\,\log (z)}{3}} \right)  \right) \Bigg]
\eeqar

\vfill
\eject

\end{document}